\documentclass[sigconf]{acmart}
\usepackage{fix-cm}
\AtBeginDocument{%
  }

\setcopyright{acmlicensed}
\copyrightyear{2025}
\acmYear{2025}
\acmDOI{XXXXXXX.XXXXXXX}
\acmConference[EASE 2025]{The 29th International Conference on Evaluation and Assessment in Software Engineering}{17–20 June, 2025}{Istanbul, Türkiye}
\acmISBN{978-1-4503-XXXX-X/2025/XX}

\usepackage{multirow}
\usepackage{booktabs} 
\usepackage{tabularray}
\usepackage{times}
\usepackage{latexsym}
\usepackage{stfloats}
\usepackage{listings}
\usepackage[listings,skins,breakable]{tcolorbox}
\usepackage{amsmath}
\colorlet{punct}{red!60!black}
\definecolor{background}{HTML}{EEEEEE}
\definecolor{delim}{RGB}{20,105,176}
\colorlet{numb}{magenta!60!black}
\lstdefinelanguage{json}{
    basicstyle=\normalfont\ttfamily,
    showstringspaces=false,
    breaklines=true,
    frame=lines,
    backgroundcolor=\color{background},
    literate=
     *{0}{{{\color{numb}0}}}{1}
      {1}{{{\color{numb}1}}}{1}
      {2}{{{\color{numb}2}}}{1}
      {3}{{{\color{numb}3}}}{1}
      {4}{{{\color{numb}4}}}{1}
      {5}{{{\color{numb}5}}}{1}
      {6}{{{\color{numb}6}}}{1}
      {7}{{{\color{numb}7}}}{1}
      {8}{{{\color{numb}8}}}{1}
      {9}{{{\color{numb}9}}}{1}
      {:}{{{\color{punct}{:}}}}{1}
      {,}{{{\color{punct}{,}}}}{1}
      {\{}{{{\color{delim}{\{}}}}{1}
      {\}}{{{\color{delim}{\}}}}}{1}
      {[}{{{\color{delim}{[}}}}{1}
      {]}{{{\color{delim}{]}}}}{1},
}
\begin{document}

\title{CallNavi, A Challenge and Empirical Study on LLM Function Calling and Routing}

\author{
 \textbf{Yewei Song\textsuperscript{1}},
 \textbf{Xunzhu Tang\textsuperscript{1}}, \textbf{Cedric Lothritz\textsuperscript{2}}
 \\
 \textbf{Saad Ezzini\textsuperscript{3,5}},
 \textbf{Jacques Klein\textsuperscript{1}},
 \textbf{Tegawendé F. Bissyandé\textsuperscript{1}}
 \\
 \textbf{Andrey Boytsov\textsuperscript{4}},
 \textbf{Ulrick Ble \textsuperscript{4}},
 \textbf{Anne Goujon\textsuperscript{4}}
\\
 \textsuperscript{1}University of Luxembourg,
 \textsuperscript{2}Luxembourg Institute of Science and Technology,
 \\
 \textsuperscript{3}Department of Information and Computer Science, KFUPM,
 \textsuperscript{4}BGL BNP PARIBAS,
 \textsuperscript{5}Interdisciplinary Research Center for Intelligent Manufacturing and Robotics, KFUPM,
% \\
%  \small{
%    \textbf{Correspondence:} \href{mailto:email@domain}{yewei.song@uni.lu}
%  }
}

\renewcommand{\shortauthors}{Trovato et al.}

\begin{abstract}
API-driven chatbot systems are increasingly integral to software engineering applications, yet their effectiveness hinges on accurately generating and executing API calls. This is particularly challenging in scenarios requiring multi-step interactions with complex parameterization and nested API dependencies. Addressing these challenges, this work contributes to the evaluation and assessment of AI-based software development through three key advancements: (1) the introduction of a novel dataset specifically designed for benchmarking API function selection, parameter generation, and nested API execution; (2) an empirical evaluation of state-of-the-art language models, analyzing their performance across varying task complexities in API function generation and parameter accuracy; and (3) a hybrid approach to API routing, combining general-purpose large language models for API selection with fine-tuned models and prompt engineering for parameter generation. These innovations significantly improve API execution in chatbot systems, offering practical methodologies for enhancing software design, testing, and operational workflows in real-world software engineering contexts.\end{abstract}





\maketitle

\section{Introduction}
\begin{figure}[htbp]
    %\centering
    \includegraphics[width=0.8\linewidth]{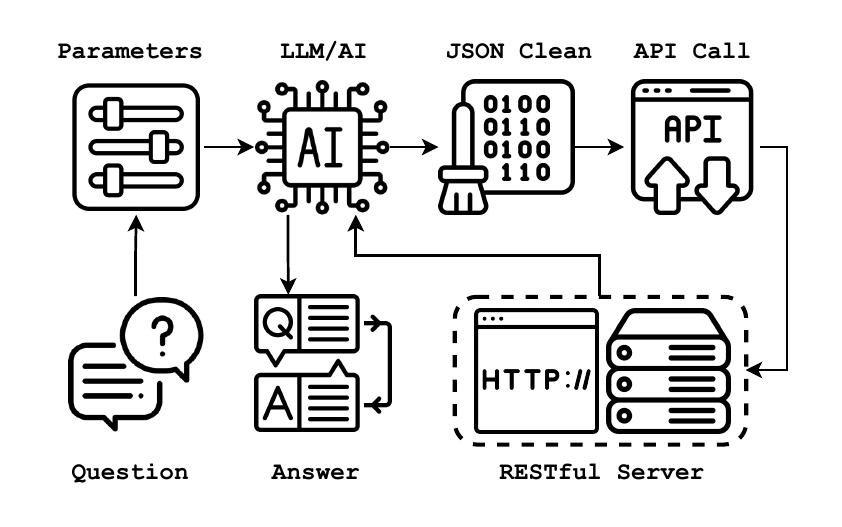}
    \caption{Example of API Calling pipeline via LLM}
    \label{fig:calling}
    \vspace{-2em}
\end{figure}

Modern conversational AI systems, such as chatbots, rely on accurate API calling to enable effective user interactions, as shown in Figure \ref{fig:calling}. Beyond generating simple API calls, models must handle complex scenarios involving selecting the correct API from extensive lists, orchestrating multiple sequential calls, and managing nested API interactions. While progress has been made in generating syntactically correct \textbf{single} API calls, there is limited focus on generating \textbf{sequences} of API calls with logical dependencies in long description context, a crucial requirement for real-world applications. 

Large Language Models (LLMs), such as GPT-4~\cite{achiam2023gpt} and Llama~\cite{touvron2023llama}, have demonstrated impressive capabilities in various natural language processing tasks. These models excel at generating coherent and contextually relevant responses, but their ability to produce structured outputs, such as API calls, program code, or other machine-readable formats, remains a challenging frontier. Structured output generation requires adherence to predefined syntactic and semantic rules, making it more constrained than generating free-form text\cite{10.1145/3613905.3650756}.

Recent advancements have explored structured output generation in applications such as code generation~\cite{chen2021evaluating}, table completion~\cite{herzig2020tapas}, and multi-turn dialogue~\cite{budzianowski2018multiwoz}. Tools like CodeX~\cite{chen2021evaluating} and AlphaCode~\cite{li2022competition} focus on generating functionally valid code, while methods like chain-of-thought prompting~\cite{wei2022chain} and tool-augmented reasoning frameworks~\cite{qin2023toolllm} enhance reasoning in complex tasks. These methods highlight the potential of LLMs for tasks requiring step-by-step reasoning and structured output generation.

Early studies, such as BotBase~\cite{zamanirad2017programming}, explored translating natural language into API calls, laying the groundwork for automating tool use. More recent benchmarks, including API-Bank~\cite{li2023api}, ToolEyes~\cite{ye2024tooleyes}, and BFCL~\cite{berkeley-function-calling-leaderboard}, evaluate LLMs on API execution. However, these datasets often operate with small API candidate pools or lack scenarios involving nested or interdependent API calls. For example, API-Bank assesses tool-augmented models but limits API candidates to fewer than five per task. Similarly, ToolBench~\cite{liu2024toolace} and ToolEyes evaluate multi-tool scenarios but do not support tasks requiring highly interdependent API calls.

To address these gaps, we propose \texttt{CallNavi}, a novel benchmark designed to evaluate LLMs on:
\begin{itemize}
    \item Selecting APIs from an unfiltered list of over 100 candidates;
    \item Executing multiple sequential API calls;
    \item Handling nested API interactions.
\end{itemize}

\texttt{CallNavi} introduces real-world complexity by simulating unfiltered API selection and combining generation with routing tasks in a long context input. It categorizes questions into easy, medium, and hard levels, enabling a granular evaluation of model capabilities across varying task complexities. Additionally, we propose new metrics, including a stability score, to measure prediction consistency across multiple runs.

We benchmark 18 LLMs, encompassing commercial, general-purpose, and fine-tuned models, on \texttt{CallNavi}. Our findings provide insights into the strengths and limitations of current models, laying a foundation for advancing API selection and function calling.
\section{Related Work}
Generating and executing accurate API calls is crucial to integrating LLM into real-world conversation applications. Existing benchmarks, such as API-Bank~\cite{li2023api}, ToolEyes~\cite{ye2024tooleyes}, and ToolBench~\cite{liu2024toolace}, evaluate API selection and execution capabilities but often rely on prefiltered API candidate pools, lack nested API tasks, or focus on narrow domain coverage. In contrast, \texttt{CallNavi} introduces unfiltered API selection with over 100 candidates, multi-call tasks, and nested API scenarios across 10 diverse domains. Table~\ref{tb:comparison} highlights these distinctions, demonstrating how \texttt{CallNavi} addresses limitations in existing benchmarks by introducing realistic complexity and structured difficulty levels.

Structured output generation, a critical capability for API function calling, has seen significant advancements. MetaGPT~\cite{hong2023metagpt} and CodeAgent~\cite{tang2024codeagentautonomouscommunicativeagents} emphasize task decomposition and multi-step reasoning, improving performance in complex workflows. Techniques like constrained generation~\cite{beurer2024guiding} and grammar-aware Seq2Seq models~\cite{dong2023codep} improve structured output reliability, aligning with \texttt{CallNavi}'s focus on evaluating structured reasoning and accuracy.

Stability in LLM predictions is another vital area of research. Although traditional metrics such as \textit{freq@topk} assess prediction reliability, they fail to capture consistency across multiple runs fully. Inspired by prior stability-focused studies~\cite{du2023shortcutlearninglargelanguage, tian2023chatgpt}, \texttt{CallNavi} introduces a stability score to quantify prediction consistency, complementing traditional metrics like AST match and exact match.

From a software engineering perspective, function-calling tasks align with modularity and abstraction principles, emphasizing decomposition into manageable sub-tasks. Early works, such as BotBase~\cite{zamanirad2017programming}, synthesized API calls from natural language, laying the groundwork for modern tools like Gorilla~\cite{gorilla-openfunctions-v2}, ToolLLM~\cite{qin2023toolllm}, and ToolAlpaca~\cite{tang2023toolalpaca}. Recent efforts like StableToolBench~\cite{guo2024stabletoolbench} and $\tau$-bench~\cite{yao2024tau} highlight challenges in tool learning and real-world tool-agent-user interactions.

API recommendation systems, such as those explored by Peng et al.~\cite{peng2022revisiting}, provide insights into ranking and selecting APIs. These systems complement \texttt{CallNavi}, which emphasizes multi-step workflows requiring careful API selection. Similarly, abstract syntax networks~\cite{rabinovich2017abstract} and benchmarks like BigCodeBench~\cite{zhuo2024bigcodebench} advance structured code generation and semantic parsing, aligning with \texttt{CallNavi}'s emphasis on reasoning and logical consistency in nested tasks.

In summary, while prior work has laid a strong foundation for API function calling, \texttt{CallNavi} advances the field by addressing critical gaps such as unfiltered API selection, nested tasks, and stability evaluation. These contributions provide a robust framework for benchmarking LLMs in realistic and complex scenarios.

% \section{Related Work}
% \subsection{Function Calling and API Selection}
% Table~\ref{tb:comparison} compares \texttt{CallNavi} with existing benchmarks across key dimensions, highlighting its unique contributions. Notably, our \texttt{CallNavi} is the only benchmark to test models on unfiltered API selection with over 100 candidates. And \texttt{CallNavi} incorporate both multi-call and nested API tasks in a single dataset, provide a diverse set of 10 domains while maintaining structured difficulty levels.

\begin{table}[h!]
\centering
\caption{Comparison of \texttt{CallNavi} with existing API function-calling benchmarks test set.}
\label{tb:comparison}
\resizebox{0.5\textwidth}{!}{%
\begin{tabular}{l|c|c|c|c|c}

\toprule
\textbf{Benchmark} & \textbf{Domains} & \textbf{Questions} & \textbf{Max API}& \textbf{Multi-Call} & \textbf{Nested} \\ 
& & & \textbf{Candidates} & & \\ \midrule
\textbf{\texttt{CallNavi}} & 10 & 729 & 115 & Yes & Yes \\ \hline
API-Bank & 8 & 753 & $<$5 & Yes & Yes \\ \hline
ToolEyes & 41 & 382 & $<$20 & Yes & No \\ \hline
ToolBench & 8 & 795 & 32 & Yes & No \\ \hline
BFCL (API) & N/A & 70 & $<$5 & Yes & No \\ \bottomrule
\end{tabular}%
}
\vspace{-2em}
\end{table}

\section{Research Questions}
\begin{itemize}
    \item \textbf{Benchmark} Which LLMs have the best performance for function calling in a real-world scenario?
    \item \textbf{Evaluation} Which is the best way to evaluate the API function calling ability of LLMs? 
    \item \textbf{Optimization} How to enhance API function calling ability for zero/few-shot LLM?
\end{itemize}
\section{CallNavi Dataset}
To create our dataset, we adopted a hybrid approach that combines automated generation with manual validation and construction to ensure high-quality and diverse data across different levels of difficulty. The process consisted of the following steps:

\paragraph{Initial API Function Generation.} 
Using GPT-4o, we generated API function names, descriptions, parameters, and return values based on a variety of scenario descriptions spanning multiple domains. These domains were selected to reflect realistic use cases across 10 common chatbot application areas, as described in Table~\ref{tb:dataset}. This ensures that the dataset evaluates \texttt{CallNavi} in scenarios requiring advanced task routing, contextualization, and regulatory compliance.

\paragraph{Validation and Refinement.}
All generated API functions were manually reviewed for accuracy, consistency, and relevance. i.e.:
\begin{itemize}
    \item Parameters were checked to ensure they aligned with real-world API design conventions.
    \item Ambiguities or redundancies in function descriptions were resolved.
    \item Naming conventions for parameters and return values were standardized to ensure consistency across the dataset.
\end{itemize}

\paragraph{Generation of Easy Questions.}
For the easy subset, we used GPT-4o to generate questions related to API usage. These questions were subsequently validated to ensure:
\begin{itemize}
    \item Relevance to the provided APIs,
    \item Syntactic and semantic correctness, and
    \item Coverage of straightforward, single API usage scenarios.
\end{itemize}

\paragraph{Manual Construction of Medium and Hard Questions.}
Medium and hard questions were manually crafted to reflect increasingly complex API calling scenarios. The criteria and considerations for these levels were as follows:
\begin{itemize}
    \item \textbf{Medium Questions:} Focused on multi-step tasks requiring the use of multiple APIs in sequence. These tasks test the model's ability to identify dependencies between API calls while maintaining logical flow.
    \item \textbf{Hard Questions:} Designed to address edge cases, ambiguous queries, and nested API calls requiring advanced reasoning. Scenarios simulate real-world challenges, such as incomplete user inputs or conflicting requirements.
\end{itemize}

\paragraph{Quality Control.}
The dataset underwent a multi-stage quality assurance process to ensure its reliability:
\begin{itemize}
    \item Each generated instance was cross-checked by multiple annotators for correctness and consistency.
    \item For manually written instances, authors verified adherence to the design criteria.
    \item Errors, ambiguities, and inconsistencies were flagged and resolved iteratively.
\end{itemize}

\paragraph{Summary.}
\label{sec:meta}
The \texttt{CallNavi} dataset combines automation with human oversight, resulting in a benchmark that is both realistic and challenging. By spanning easy, medium, and hard tasks across diverse real-world domains, as outlined in Table~\ref{tb:dataset}, the dataset evaluates LLM capabilities in scenarios requiring robust task routing, contextual understanding, and API management. 

The first part of the metadata is a long JSON file with the API name, description, and parameters in the following format.
\begin{lstlisting}[language=json,firstnumber=1]
{
    "name": "getAccountBalance",
    "parameters": ["accountID"],
    "description": "Retrieves the current balance for a specific account.",
    "returnParameter": {
      "Balance": "number"
    }
},
...
\end{lstlisting}
We then format each question as shown in the example below, which includes the user query, the ground truth API call in JSON format, and the difficulty level:
% \cl{sentence needs to be revised}
\begin{lstlisting}[language=json,firstnumber=1]
{
  "id": "ban01",
  "question": [
    {"role": "user", 
    "content": "What is the balance for the account with ID 987654?"}],
  "ground_truth": {
    "API": ["getAccountBalance"],
    "parameters": 
        {"accountID": "987654"}},
  "difficulty": "easy"
},
...
\end{lstlisting}
\begin{table}[h!]
\centering 
\caption{CallNavi dataset domains, questions and difficulties statistics table.}
\label{tb:dataset}
\resizebox{0.47\textwidth}{!}{
\begin{tblr}{
  cell{1}{1} = {r=2}{},
  cell{1}{2} = {r=2}{},
  cell{1}{3} = {r=2}{},
  cell{1}{4} = {c=3}{c},
  cell{1}{7} = {r=2}{},
  hline{1,3-14} = {-}{},
  hline{2} = {4-6}{},
  column{7}  = {r}
}
% \toprule
Domain             & {API\\Functions} & Questions & Difficulty     &                   &       & {Max Input\\Tokens}         \\
                   &           &           & Easy  & Medium  & Hard &  \\
Banking            & 91        & 115       & 70             & 28                & 17      &    6517    \\
Shopping           & 81        & 65        & 41             & 17                & 7    &   5195        \\
Logistics          & 46        & 65        & 40             & 17                & 8    &      3434     \\
Aviation           & 48        & 80        & 44             & 24                & 12      &   3461     \\
Healthcare         & 20        & 47        & 31             & 10                & 6    &    1788       \\
Public Services    & 82        & 85        & 50             & 27                & 8        &   6249    \\
Human Resources    & 20        & 35        & 21             & 13                & 1        &  1863     \\
Hotel Industry     & 49        & 65        & 40             & 19                & 6           & 3811   \\
Insurance          & 42        & 60        & 40             & 11                & 9          &  3452   \\
Telecommunications & 100       & 112       & 79             & 22                & 11          & 6374   \\
Overall            & 579       & 729       & 456            & 188               & 85       &  6517    
\end{tblr}
}
\end{table}

\subsection{Dataset}
\footnote{https://github.com/Etamin/CallNavi}
The \textit{CallNavi} dataset evaluates LLMs' task routing and API calling capabilities across multiple domains. As shown in Table~\ref{tb:dataset}, it contains \textbf{729 questions} of varying difficulty and API interaction complexity, along with \textbf{579 distinct API functions}. Questions are categorized into \textbf{easy}, \textbf{medium}, and \textbf{hard} levels:

\begin{itemize}
    \item \textbf{Easy(456 questions)}: Require a \textbf{single API} call to fulfill task.\\
    \textit{Example}: A user checking their bank account balance with one straightforward API call.

    \item \textbf{Medium(188 questions)}: Involve \textbf{multiple APIs} within the same question, with all parameters provided in the context.\\
    \textit{Example}: A shopping query needing product details and stock availability via two independent API calls.

    \item \textbf{Hard(85 questions)}: Require \textbf{multiple API calls} where some parameters depend on \textbf{responses from previous calls}, adding complexity. \textbf{5} steps maximum of API.\\
    \textit{Example}: Updating delivery status by first retrieving a package ID, then using it to fetch the delivery status through sequential API calls.
\end{itemize}

This dataset tests LLMs' ability to perform function-calling routing and parameters generation across varying difficulties, assessing both basic single-call handling and complex multi-step nested requests. Zero-shot and few-shot models must infer correct API interactions only from the question context from different difficulties.

\section{Metrics and Evaluation}
\subsection{API Parameters AST Match}
\label{sec:score}
Our study utilizes Abstract Syntax Tree (AST) evaluation to assess models' ability to generate accurate JSON outputs for API calls. The format of the output JSON and parameters follows a structure similar to the BFCL dataset in Section \ref{sec:meta} ~\cite{berkeley-function-calling-leaderboard}. We parse the generated JSON string into an object and compare each component with the ground truth, such as the API list and parameters.

In scenarios involving multi-step API calls where parameters depend on previous steps or where text inputs may not have a single definitive answer, placeholder tokens are used for parameters. These tokens are positionally aligned with the ground truth, and we exclude them from strict comparisons during evaluation.

\begin{figure}[htbp]
    %\centering
    \hspace{-0.5cm}
    
    \includegraphics[width=1\linewidth]{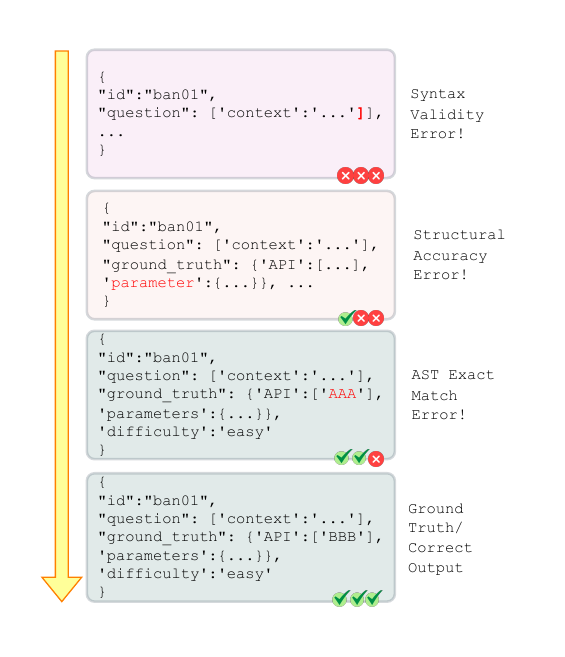}
    \caption{Example of Evaluation Pipeline}
    \label{fig:exammple}
    \vspace{-1em}
\end{figure}

\begin{figure}
    %\centering
    % \hspace{-.6cm}
    \includegraphics[width=1\linewidth]{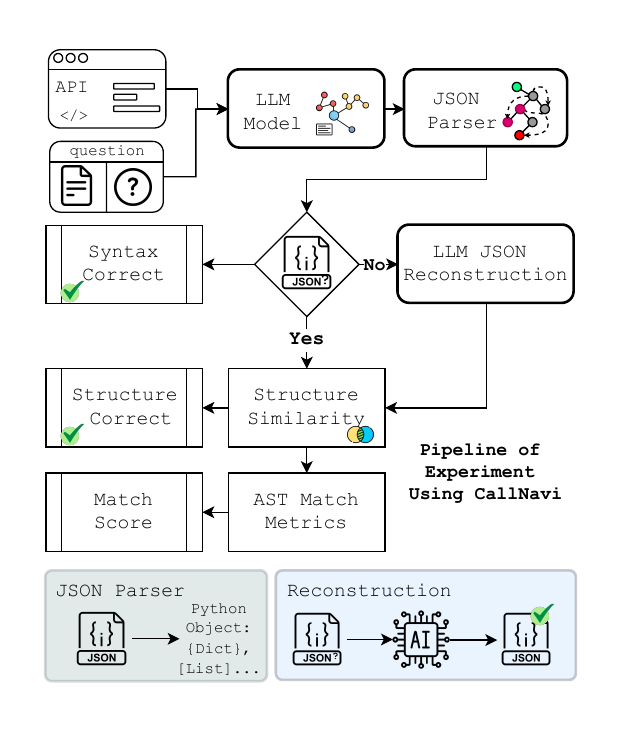}
    \caption{Pipeline of AST Match Score}
    \label{fig:pipeline}
    \vspace{-2em}
\end{figure}

Our AST evaluation process is based on three key criteria, examples in Figure \ref{fig:exammple}: \begin{itemize} 
    \item \textbf{Syntax Validity}: Whether the JSON string can be correctly parsed into an object without syntax errors. 
    \item \textbf{Structural Accuracy}: Whether the parsed API calls match the ground truth and include the correct parameter names(keys). 
    \item \textbf{AST Exact Match}: Whether the entire parsed object, including its structure and content, is identical to the ground truth.
\end{itemize}
\vspace{-1em}
As Figure \ref{fig:pipeline} shows, we begin by checking the syntax validity of the generated JSON structure. If syntax errors are detected, we apply a JSON fix prompt to repair the structure or convert alternative formats (such as function-calling syntax) into valid JSON. Once the structure is valid, we assess structural accuracy by comparing the predicted JSON with the ground truth. A structural match is scored as 1.
Finally, we convert both the predicted and ground truth JSONs into object trees, comparing each node and leaf. A perfect match across all nodes results in a score of 1 for AST Exact Match.

This multistep evaluation ensures a thorough assessment of the accuracy of API function calls and the structural integrity of the parameters, allowing for a granular analysis of the performance.
\subsection{LLM-as-a-Judge Evaluation}
\label{sec:gptscore}
We also use GPT-4o language models to evaluate whether the generated JSON outputs correspond accurately to the ground truth~\cite{zheng2023judging}. 
This approach aims to observe if LLMs can perform such evaluation tasks with high precision. 
Using an LLM for this purpose, we assess its ability to compare and validate structured data, thereby determining its effectiveness in automating the evaluation process.

\subsection{Stability Score}
\label{sec:stab_score}

In chatbot systems, consistent outputs for identical inputs are crucial to ensure reliability and user trust, especially in professional settings. Users expect the same accurate response each time they ask the same question. Inconsistencies can cause confusion, erode confidence, and lead to errors, particularly in critical fields like finance or healthcare.

We propose an \textbf{Election Stability Score} to evaluate the consistency of API outputs across multiple runs for the same input. This score mirrors an election process, selecting the majority output as the final answer. To calculate the score, we define:

\begin{itemize}
    \item $N$: Total number of outputs (samples).
    \item $F_1$: Maximum frequency among all unique outputs (count of the most frequent output).
    \item $F_2$: Second maximum frequency (count of the second most frequent output).
\end{itemize}

The stability score is calculated as:

\[
\text{Stability Score} = \frac{F_1 - F_2}{N - F_2}
\]

This quantifies the consistency of the model's outputs: If there's a tie for the most frequent output ($F_1 = F_2$), the stability score is set to $0$, indicating no consensus; If the most frequent output is unique ($F_1 > F_2$), the score ranges from $0$ to $1$, reflecting the dominance of the most frequent output.

To ensure reliable comparisons and reduce errors, we preprocess the outputs by removing unnecessary spaces, newlines, and formatting inconsistencies, converting the text to lowercase, and stripping extraneous characters that could cause mismatches.

While `freq@topk` is often used to evaluate the performance of model stability, it does not capture the stability in LLM output. For example, if a model produces the sequence "AABBC" across multiple runs, `freq@topk` might assign a high score of 0.4 because the most frequent token ("A") appears 40\% of the time. However, this sequence is unstable as \textbf{no single output} consistently dominates. In contrast, our stability score focuses on the dominance of the most frequent output, offering a better measure of a model's reliability in structured tasks.

To give a clear example of calculating the stability of the model's outputs, we analyze the frequency distribution of the results obtained from 5 times runs. Let's review the variable settings:

\begin{itemize}
    \item \( N \) be the total number of outputs (samples).
    \item \( F_1 \) be the \textbf{maximum frequency} of any unique output (the most frequent output).
    \item \( F_2 \) be the \textbf{second maximum frequency} (the frequency of the second most frequent output).
\end{itemize}

\textbf{Explanation:}

\begin{itemize}
    \item \textbf{Tie Situations (\( F_1 = F_2 \))}: When the maximum frequency is equal to the second maximum frequency, it indicates a tie for the most frequent output. The stability score is set to 0 to reflect neither majority nor consensus in such cases.
    
    \item \textbf{No Tie Situations (\( F_1 > F_2 \))}: The numerator \( F_1 - F_2 \) measures the dominance of the most frequent output over the second most frequent one. The denominator \( N - F_2 \) normalizes this difference relative to the total number of outputs excluding those of the second most frequent output. The resulting score ranges from 0 to 1; higher values indicate greater stability.
\end{itemize}

\textbf{Examples:}

\begin{itemize}
    \item \textbf{All Outputs Identical:}
    
    \begin{itemize}
        \item \textbf{Results}: All outputs are the same (e.g., \( ['A', 'A', 'A', 'A', 'A'] \)).
        \item \( F_1 = N \), \( F_2 = 0 \) (since there's only one unique output).
        \item \textbf{Stability Score}: 
        \[
        \text{Stability Score} = \frac{N - 0}{N - 0} = \frac{N}{N} = 1
        \]
        Indicates perfect stability.
    \end{itemize}

    \item \textbf{Tie Situation (e.g., 2 vs 2 vs 1):}
    
    \begin{itemize}
        \item \textbf{Results}: Two outputs occur twice, and one occurs once (e.g., \( ['A', 'A', 'B', 'B', 'C'] \)).
        \item \( F_1 = 2 \), \( F_2 = 2 \) (tie between 'A' and 'B').
        \item \textbf{Stability Score}:
        \[
        \text{Stability Score} = 0
        \]
        Reflects the lack of consensus due to the tie.
    \end{itemize}
    
    \item \textbf{Minority Advantage (e.g., 2 vs 1 vs 1 vs 1):}
    \begin{itemize}
        \item \textbf{Results}: One output occurs two times, another occurs once each (e.g., \( ['A', 'A', 'B', 'C', 'D'] \)).
        \item \( F_1 = 2 \), \( F_2 = 1 \).
        \item \textbf{Stability Score}:
        \[
        \text{Stability Score} = \frac{2 - 1}{5 - 1} = \frac{1}{4} \approx 0.25
        \]
        Indicates a little stability.
    \end{itemize}
    
    \item \textbf{Partial Agreement(Strong Opposition) (e.g., 3 vs 2):}
    
    \begin{itemize}
        \item \textbf{Results}: One output occurs three times, another occurs twice (e.g., \( ['A', 'A', 'A', 'B', 'B'] \)).
        \item \( F_1 = 3 \), \( F_2 = 2 \).
        \item \textbf{Stability Score}:
        \[
        \text{Stability Score} = \frac{3 - 2}{5 - 2} = \frac{1}{3} \approx 0.333
        \]
        Indicates moderate stability.
    \end{itemize}

    \item \textbf{Partial Agreement(Weak Opposition) (e.g., 3 vs 1 vs 1):}
    
    \begin{itemize}
        \item \textbf{Results}: One output occurs three times, another occurs once each (e.g., \( ['A', 'A', 'A', 'B', 'C'] \)).
        \item \( F_1 = 3 \), \( F_2 = 1 \).
        \item \textbf{Stability Score}:
        \[
        \text{Stability Score} = \frac{3 - 1}{5 - 1} = \frac{1}{2} \approx 0.5
        \]
        Indicates higher moderate stability.
    \end{itemize}
    
    \item \textbf{High Majority (e.g., 4 vs 1):}
    
    \begin{itemize}
        \item \textbf{Results}: One output occurs four times, another occurs once (e.g., \( ['A', 'A', 'A', 'A', 'B'] \)).
        \item \( F_1 = 4 \), \( F_2 = 1 \).
        \item \textbf{Stability Score}:
        \[
        \text{Stability Score} = \frac{4 - 1}{5 - 1} = \frac{3}{4} = 0.75
        \]
        Indicates strong stability.
    \end{itemize}
    
    \item \textbf{All Outputs Unique:}
    
    \begin{itemize}
        \item \textbf{Results}: All unique outputs (e.g., \( ['A', 'B', 'C', 'D', 'E'] \)).
        \item \( F_1 = F_2 = 1 \).
        \item \textbf{Stability Score}:
        \[
        \text{Stability Score} = 0
        \]
        Complete instability due to a lack of consensus.
    \end{itemize}
\end{itemize}

\textbf{Interpretation:}

\begin{itemize}
    \item \textbf{Stability Score of 1}: Perfect stability; all outputs are same.
    \item \textbf{Stability Score of 0}: No stability; either all outputs are unique, or there's a tie for the most frequent output.
    \item \textbf{Stability Scores Between 0 and 1}: Partial stability; higher scores indicate greater agreement among outputs.
\end{itemize}

To further evaluate the stability of the model's outputs, we calculate the average Levenshtein distance between the first answer and each subsequent output\cite{zhang2017research}. We normalize Levenshtein distance using the following formula. 

\[
Score_{Lev} = \frac{1}{n} \sum_{i=1}^{n}(1-\frac{lev(x_0,x_i)}{Max(len(x_0),len(x_i))} ) 
\]

% \cl{Levenshtein distance is well-known. Feel free to remove the explanation and formula}
\section{Experiments and Benchmark}

\subsection{Models}
To evaluate the benchmark, we selected models based on their performance, architecture, and relevance to function-calling tasks. The selection criteria focused on general-purpose and fine-tuned models optimized for function calling or JSON generation, ensuring a well-rounded comparison between zero-shot and fine-tuned capabilities. Models like BART and traditional retrieval-based approaches were excluded as they lack the ability to select APIs from extensive lists, which is critical for the complexity of this task.

Table~\ref{tab:model-comparison} organizes the selected models into four groups: commercial models, medium-large models (10B + parameters), small models (5B-10B parameters) and light models (with parameters below 5B). 
% Commercial models like GPT-4o and Gemini 1.5 Flash represent state-of-the-art performance in general-purpose tasks, while middle-large models, such as LLaMA 3.1 (70B) and Command-R (35B), cater to more resource-intensive, complex tasks. Small models, including Gemma2 (9B) and LLaMA 3.1 (8B), balance efficiency and performance, while lightweight models like NemoTron-Mini (4B) and Phi3.5 (3B) are suitable for resource-constrained applications. 

\begin{table}[h!]
\centering
\resizebox{0.45\textwidth}{!}{
\begin{tabular}{llrr}
\toprule
\textbf{Model Name} & \textbf{Origin} & \textbf{Size} & \textbf{Context Limits} \\ \hline
GPT-4o & OpenAI~\cite{achiam2023gpt} & N/A & 128K \\
GPT-4o-mini & OpenAI~\cite{achiam2023gpt} & N/A & 128K \\ 
Gemini 1.5 Flash & Google~\cite{reid2024gemini} & N/A & 1M \\\hline
LLaMA 3.1 & Meta AI~\cite{meta-ai-2024} & 70B & 128K\\
Command-R & Command AI~\cite{commandr} & 35B & 128K\\
Gemma2 & Google~\cite{team2024gemma} & 27B & 8K\\
Mistral-Small & Mistral AI~\cite{mistralai-2024} & 22B & 128K\\
Phi3 & Microsoft~\cite{abdin2024phi} & 14B &128K\\
Mistral-Nemo & Mistral AI~\cite{mistralnemo-2024} & 12B & 1M\\ \hline
Gemma2 & Google~\cite{team2024gemma} & 9B & 8K\\
LLaMA 3.1 & Meta AI~\cite{meta-ai-2024} & 8B & 128K\\ 
xLAM & Salesforce~\cite{zhang2024xlam} & 7B & 4K\\ 
DeepSeek R1 & DeepSeek-AI~\cite{guo2025deepseek} & 7B & 128K\\ \hline
NemoTron-Mini & NVIDIA~\cite{paul-2024} & 4B & 4K \\
Phi3.5 & Microsoft~\cite{trufinescu-2024} & 3B & 128K \\
LLaMA 3.2 & Meta AI~\cite{llama3.2-2024} & 3B & 128K\\ \hline
NexusRaven & Nexusflow.ai~\cite{nexusraven} & 13B &16K \\
Gorilla & Berkeley~\cite{patil2023gorilla} & 7B & 4K\\
\bottomrule
\end{tabular}}
\caption{Comparison of General Purpose LLMs.}
\label{tab:model-comparison}
\vspace{-2em}
\end{table}

We chose 2 fine-tuned Function Calling models for testing, which have top performance on the BFCL leaderboard: NexusRaven and Gorilla OpenFunctions v2.  Then we found that some models cannot input long lists, e.g. Firefunction v2~\cite{Garbacki_Chen_2024}. 

\begin{table*}[htb]
\centering

\resizebox{0.9\textwidth}{!}{\begin{tblr}{
  row{1} = {c},
  row{2} = {c},
  cell{1}{1} = {r=2}{},
  cell{1}{2} = {r=2}{},
  cell{1}{3} = {c=4}{},
  cell{1}{7} = {r=2}{},
  cell{1}{8} = {r=2}{},
  cell{1}{9} = {c=5}{},
  cell{1}{14} = {r=2}{},
  cell{3}{1} = {r=3}{},
  cell{6}{1} = {r=6}{},
  cell{12}{1} = {r=4}{},
  cell{16}{1} = {r=3}{},
  cell{19}{1} = {r=2}{},
  cell{21}{1} = {r=3}{},
  hline{1,24} = {-}{0.08em},
  hline{2} = {3-6,9-13}{},
  hline{3} = {-}{},
}
Category               & Models        &  {API Calling Routing\\Exact Match} &        &      &       & {Syntax \\Validity} & {Structural \\Accuracy} & {API Calling with Parameters\\AST Match} &        &    &  &     & {Overall\\GPT Score} \\
                       &                           & Easy                          & Medium & Hard & All &       &               &                           Easy                           & Medium & Hard & All Avg. &   Macro Avg.  &                 \\
{Commercial\\Models}                 & GPT4o         &    0.978        &       0.914                        &     0.611   &  \textbf{0.919}    &                  0.993        &      \textbf{0.887}                    &                           0.802     &     0.638   &  0.388   &     \textbf{0.711}    &   \textbf{0.609} &   \textbf{0.913}             \\
                       & GPT4o mini     & 0.971        &     0.930               &       0.564    &     0.913              &         \textbf{0.994}             &             0.869             &              0.800                  &  0.648      &   0.364   &  0.710   &  0.604 &  0.908                 \\  
                       & Gemini 1.5 Flash      &  0.973       &      0.904              &     0.564      &            0.908       &      0.945               &        0.806                &                     0.728      & 0.462       &  0.258    &  0.604   &  0.483  &    0.876          \\  \hline
{Large~\\General\\LLMs} & LLAMA3.1\:70B        &     0.945                          &    0.835    &    0.470  &  0.861   &    0.967                  &    \textbf{0.299}                      &                     0.296           &   0.191     &  0.094    &  \textbf{0.245}   &   \textbf{0.194}    &    \textbf{0.583}           \\

                       & CommandR\:35B        &                    0.789          &     0.877     &  0.529    & 0.781    &        0.969              &           0.189               &   0.167                             & 0.095       &   0.047   &   0.134  &     0.103   &  0.400            \\
                       & Gemma2\:27B        &           0.945                   &    0.877   &   0.552   &  \textbf{0.882}   &      0.982                &     0.226                     &            0.217                    &  0.143      &   0.070   & 0.181  & 0.143  &  0.476                    \\  & Mistral-Small\:22B         &        0.885                       &   0.819     &   0.494   &   0.823 &    \textbf{0.986}                  &   0.196                       &       0.201                         &  0.106      &   0.059   &   0.160     & 0.122 &    0.417                  \\   & Phi3\:14B        &         0.050                      &  0.032      &  0.011        &   0.041    &  0.283   &    0.021                  &        0.019                  &    0.010                         &  0.0    &  0.015   &  0.010  &  0.082               \\  & Mistral-Nemo\:12B        &             0.927                 &     0.808   &   0.470   & 0.843    &   0.842                  &        0.271                  & 0.296                               &     0.127   & 0.035     &   0.222  &   0.153  & 0.524                \\ \hline {Middle~\\LLMs}
                       & Gemma2\:9B         &          0.962                     &    0.845    &  0.506    & \textbf{0.879}    &     0.983                &            0.220              &   0.241                             &    0.095    &    0.059  &  0.182   &  0.132  &  0.488                \\
                     & LLAMA3.1\:8B        &                     0.916          &     0.813   &    0.552  & 0.847    &    0.925                  &            0.207              &       0.223                         &   0.058     &  0.059    &   0.162  &   0.113     &  0.422            \\
                       
      & xLAM-fc\:7B         &  0.642                             &  0.377      &   0.188   &   0.521  &   \textbf{0.990}                   &             \textbf{0.271}             &      0.307                          &     0.117   &  0.058    & \textbf{0.229}    & \textbf{0.161} &   \textbf{0.554}                     \\   & DeepSeek R1\:7B         &  0.250                             &  0.271      &   0.082   &   0.235  &   0.902                   &             0.117             &        0.129   &  0.042    & 0.047    & 0.097 & 0.073   &         0.289            \\ \hline
{Light\\Models}      & nemotron-mini\:4B         &    0.644                           &  0.287      &  0.094    &   0.488  &  0.529                    &               0.080           &         0.067                       &   0.010     &  0.012    &   0.047  & 0.030  & 0.271                   \\
                       & LLAMA3.2\:3B         &    0.842                           &   0.622     &   0.400   &   \textbf{0.733}  &            \textbf{0.917}          &    0.063                      &       0.052                         &   0.021     & 0.035     &\textbf{0.042}     &       \textbf{0.036}    &  \textbf{0.353}        \\
                       & Phi3.5\:3B         &        0.723                       &    0.340    & 0.188     &  0.562   &      0.004                &           0.0               &       0.0                         &   0.0     &   0.0    &     0.0 &  0.0 & 0.002                  \\ \hline
Fine-Tuned             & NexusRaven\:13B
  &            0.210                   &   0.148     &  0.082    &   0.179  &       N/A               &    N/A                      &  0.160                              &    0.074    & 0.047    & 0.124    &    0.094    &   0.254          \\
& Gorilla v2\:7B        &      0.616                         &    0.005    &   0.0   &   \textbf{0.387}  &  N/A                    &        N/A                  &       0.524                         &   0.005     &  0.0    & \textbf{0.329}    &  \textbf{0.176}  & \textbf{0.518}                          \\ \hline 
% {Asynchronous\\Generation} & *NexusRaven\:13B
%   &            N/A                   &   N/A     &  N/A    &   N/A  &       N/A               &    N/A                      &              0.657                  &     0.457   &  0.188   &    0.551 &       0.434    &  0.795         \\
% & *Gorilla v2\:7B        &     N/A                         &    N/A    &  N/A   &   N/A  &  N/A                    &        N/A                  &     0.682                           &     0.005   &   0.0   &  0.427   &  0.229  & 0.626   \\
%                        & *xLAM-fc\:7B      &     N/A                          &   N/A     &  N/A    &   N/A  &  N/A                    &             0.777            &    0.714                            &  0.462      &    0.188  &  \textbf{0.588}   &        \textbf{0.455}   &     \textbf{0.835}
\end{tblr}}
\caption{Benchmark Results Table. Macro Avg. means the arithmetic mean of 3 difficulties.}
\label{tb:results}
\vspace{-2em}
\end{table*}
\subsection{Environment}
All our local models run with 4-bit Quantization, running on the default Ollama platform settings without any optimization for JSON generation. We do our experiments on NVIDIA-V100 GPU.
\subsection{Pipeline}
Our evaluation pipeline begins with the creation of prompts based on two templates. The first template focuses on retrieving the \textbf{API calling list}, which corresponds to the "API" list in the ground truth. This prompt instructs the model to identify which API calls should be used and in what order. The second template is designed to generate the full \textbf{API calling JSON}, including the parameters for each call.

Once the prompts are generated, we run them through each model to obtain predictions. In the first part of the evaluation, where API calls are generated without parameters, we directly calculate the exact match between the predicted API list and the ground truth and make them called \textbf{API Calling Routing}. For the second part, where full JSON outputs are provided, the results are evaluated using the three AST-related scores outlined in Section \ref{sec:score}: \textbf{Syntax Validity}, \textbf{Structural Accuracy}, and \textbf{AST Exact Match}.

Finally, we employ an LLM-as-a-judge approach, using GPT-4o to calculate a score for comparison, providing a final measure of how well the model's outputs align with the ground truth. This multi-step process ensures comprehensive evaluation across various levels of output complexity.

\subsection{Benchmarks Results}
The results presented in Table \ref{tb:results} highlight the performance of various models in different aspects of API function calling, including API calling routing accuracy, syntax validity, structural accuracy, and API parameter match through AST evaluation. OpenAI's models, GPT4o and GPT4o mini, consistently outperform the others, particularly in syntax validity (0.993 and 0.994, respectively) and overall GPT score (0.913 and 0.908). Both models also demonstrate strong structural accuracy and API parameter AST match, especially in easier tasks. Gemini 1.5 Flash follows these metrics closely.

Among the large general-purpose open LLMs, LLAMA3.1 (70B) performs well in API calling with an exact match score of 0.945 in easy tasks, though its performance drops significantly in harder cases (0.470). It also achieves the second-highest overall GPT score (0.583), largely due to high syntax validity (0.967). However, its structural accuracy and parameter AST match are weaker, with significant drops in harder tasks. But middle-size LLMs show strong potential ability, very close to the larger group performance such as Gemma2(9B) and xLAM(7B).

The other models and fine-tuned models generally struggle across all indicators. For example, CommandR (35B) shows relatively strong performance in medium API calling tasks (0.877) but performs poorly in structural accuracy (0.189) and API parameter AST match (0.134). Similarly, Mistal models show moderate performance, but the smaller models (e.g., Phi3, LLAMA3.2) display particularly low overall GPT scores and poor performance in most tasks.

Our analysis demonstrates that the Pearson correlation between the "GPT Score" and the "All Avg." column in the "Parameter AST Match" section is \textbf{0.934}, with a p-value of \textbf{4.40e-08}. This indicates a very strong positive correlation, suggesting that higher GPT scores are closely associated with better average AST match performance. The results in our table are closely aligned with those of the Berkeley Function Calling Leaderboard\footnote{https://gorilla.cs.berkeley.edu/leaderboard.html}, which assesses LLMs' performance in API or function-calling tasks. In both evaluations, models like OpenAI’s GPT4o stand out for their high syntax validity and overall accuracy, as reflected in our table where GPT4o scores above 0.9 in both categories. This matches the leaderboard’s top models, which also excel in AST evaluations and execution accuracy. In contrast, fine-tuned models such as FireFunction V2 in our results show weaker performance in API calling accuracy and AST matching, a trend similarly observed with fine-tuned models like Gorilla OpenFunctions on the Berkeley leaderboard, particularly in more complex API scenarios. Both evaluations emphasize the challenges faced by fine-tuned models in handling complex function scenarios or multi-step API calls, highlighting the need for improvement in these areas.

\begin{tcolorbox}[leftrule=0mm,rightrule=0mm,toprule=0mm,bottomrule=0mm,left=0pt,right=0pt,top=0pt,bottom=0pt,title={RQ1: Which LLM have the best performance for function calling in a reality scenario?}]
    \textbf{Answer:} Based on the results of our benchmark, we can still claim that OpenAI GPT models are the best solution to solve this kind of challenge. But we can see if the test only by calling the API name list, open-source models can have a closer performance to the state-of-the-art.
\end{tcolorbox}
\subsubsection{Stability Test}

In our stability experiments, we ran 5 times referring to the previous study~\cite{tian2023chatgpt}, and the stability results are in Table~\ref{tbl:stab}. By these metrics results, we obtain a numerical comparison that reflects the stability differences between outputs. 

This stability score provides a quantitative measure of the consistency of the model's outputs across multiple runs. It accounts for both the dominance of the most frequent output and the impact of significant minority outputs, offering a nuanced assessment of model stability.

 % a larger average distance indicates greater differences and, consequently, worse stability. This metric serves as a complementary assessment to our previously defined stability score based on frequency distributions. By comparing both the Levenshtein-based score and the frequency-based stability score, we gain a more comprehensive understanding of the model's consistency and reliability in generating accurate and stable outputs.

As mentioned in Section~\ref{sec:stab_score}, a higher Election Stability Score indicates greater absolute consistency in the model's outputs across multiple runs. A high  Levenshtein Stability Score means similar between the same input in text generation output. \textbf{Commercial models} also perform better in Table \ref{tbl:stab} below.
\begin{tcolorbox}[leftrule=0mm,rightrule=0mm,toprule=0mm,bottomrule=0mm,left=0pt,right=0pt,top=0pt,bottom=0pt,title={RQ2: Which way is the best way to evaluate the API function calling ability of LLMs?}]
    \textbf{Answer: }The best way to evaluate the API function calling ability of LLMs is still \textbf{AST evaluation} with parameters. However, our \textbf{Election Stability Score} provides additional insights into output stability, revealing differences that traditional metrics may overlook. 
\end{tcolorbox}
\begin{table}[htbp]
\centering
\caption{Stability Test Results}
\label{tbl:stab}
\resizebox{0.45\textwidth}{!}{
\begin{tabular}{lrrr} 
\toprule
Model         & Size & \multicolumn{1}{c}{\begin{tabular}[c]{@{}c@{}}Election \\Stability Score\end{tabular}} & \multicolumn{1}{c}{\begin{tabular}[c]{@{}c@{}} Levenshtein \\Stability Score\end{tabular}}  \\ \hline
GPT4o         & N/A  &           0.674                          &             0.972                                                                            \\
GPT4o mini    & N/A  &              0.855                       &                                        \textbf{0.984}                                                 \\
Gemini 1.5 Flash    & N/A  &              0.825                       &                                        0.946                                                 \\
LLAMA3.1      & 70B  &                   0.407                  &                                         0.841                                                \\
LLAMA3.1      & 8B   &                    0.332                 &                                       0.740                                                  \\
Mistral-Small & 22B  &            0.208                         &                                 0.719                                                        \\
Mistral-Nemo  & 12B  &             0.365                           &                               0.734                                                          \\
CommandR      & 35B  &              0.325                       &                                      0.754                                                   \\
Gemma2        & 27B  &                0.609                     &                                        0.890                                                 \\
Gemma2        & 9B   &                    0.355                 &                                            0.864                                             \\
nemotron-mini & 4B   &                     0.013                &                                        0.527                                                 \\
LLAMA3.2      & 3B   &                   0.085                  &                            0.613                                                             \\
Phi3.5              &   3B   &        \textbf{0.909}                             &       0.637                                                                                  \\
xLAM-fc  &  7B & 0.782 &  0.948\\
DeepSeek R1  &  7B & 0.058 &  0.501\\
\bottomrule
\end{tabular}
}
\vspace{-1em}
\end{table}

\section{Zero Shot Improvement}
\subsection{Calling + Parameters 2 Steps Generation}
\label{sec:asyn}
% We observed that most general-purpose LLM perform better when generating only the API calling names compared to generating both the calling names and parameters simultaneously, as presented in Table \ref{tb:results}. This indicates that the additional complexity of producing detailed parameters alongside API calls can negatively impact the models' overall performance. Moreover, fine-tuned models struggle to handle long lists of APIs, which limits their effectiveness in scenarios requiring multiple API calls.

% To overcome these challenges, we propose a new approach that combines the strengths of general-purpose LLMs and fine-tuned models. Specifically, we utilize a general LLM to select the relevant APIs based on the input prompt, leveraging its superior capability in understanding and identifying appropriate API calls. We then provide only these selected APIs to a fine-tuned model, which focuses on generating the correct API calls along with the necessary parameters. This asynchronous generation process allows the general LLM to efficiently handle API selection, while the fine-tuned model concentrates on accurately producing the API calls and their parameters within a more manageable context.

% The effectiveness of this combined approach is demonstrated in Table \ref{tb:results}, models using this method are marked with an asterisk (*) before their names. The results show significantly improved performance, highlighting that separating the tasks of API selection and parameter generation can enhance the models' ability to handle complex API calling tasks more effectively.

We observed that most general-purpose LLMs perform better when generating only API names(routing) rather than both names and parameters simultaneously (see Table~\ref{tb:results}). The added complexity of producing detailed parameters alongside API calls can negatively impact overall performance. Additionally, fine-tuned models struggle with long lists of APIs, limiting their effectiveness in scenarios requiring multiple API calls.
\begin{figure}
    %\centering
    % \hspace{-.6cm}
    \includegraphics[width=1\linewidth]{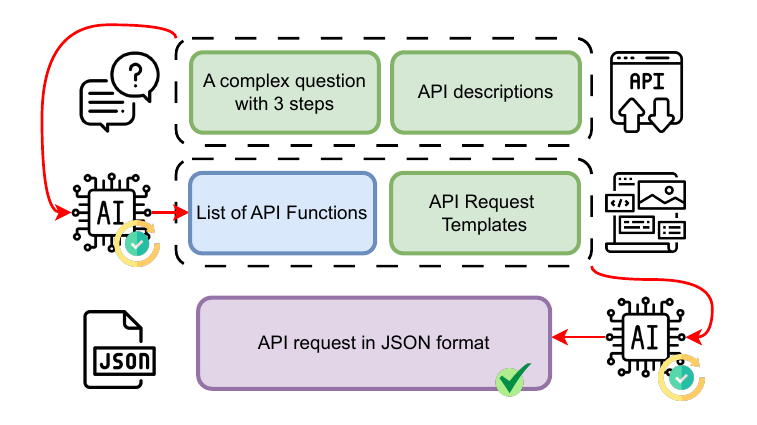}
    \caption{2-Steps Generation Pipeline}
    \label{fig:asyn}
    \vspace{-1em}
\end{figure}
To address these challenges, we propose combining the strengths of general-purpose LLMs and fine-tuned models shown in Figure\ref{fig:asyn}. Specifically, a general LLM selects the relevant APIs based on the input prompt, leveraging its superior understanding in identifying appropriate API calls. These selected APIs are then provided to a fine-tuned/LAM model, which focuses on generating the correct API calls along with the necessary parameters. This sequential process allows the general LLM to efficiently handle API selection, while the fine-tuned model concentrates on accurately producing API calls and parameters within a more manageable context.

As demonstrated in Table~\ref{tab:asyn}, this combined approach with \textbf{GPT-4o routing} significantly improves performance. Separating the tasks of API selection and parameter generation enhances the models' ability to handle complex API calling tasks more effectively.

\begin{table}[htbp]
\caption{2 Steps Generation results for LLMs}
\label{tab:asyn}
\resizebox{\columnwidth}{!}{%
\begin{tabular}{llllll}
\hline
 & Models        & easy  & medium & hard  & overall \\ \hline
\multirow{3}{*}{\begin{tabular}[c]{@{}l@{}}Fine-Tuned\\ Model w/\\ GPT routing\end{tabular}}                  & NexusRaven13B & 0.657 & 0.457 & 0.188 & 0.551 \\
 & Gorilla v27B  & 0.682 & 0.005  & 0.000 & 0.427   \\
 & xLAM-fc-7B    & 0.714 & 0.462  & 0.188 & 0.588   \\ \hline
\multirow{9}{*}{\begin{tabular}[c]{@{}l@{}}General\\ Large\\ Language\\ Models w/\\ GPT routing\end{tabular}} & Gemma2:27b    & 0.633 & 0.617 & 0.341 & 0.595 \\
 & Gemma2        & 0.723 & 0.457  & 0.164 & 0.589   \\
 & llama3.1      & 0.714 & 0.457  & 0.164 & 0.584   \\
 & mistral-small & 0.728 & 0.436  & 0.294 & 0.602   \\
 & mistral-nemo  & 0.712 & 0.308  & 0.141 & 0.541   \\
 & phi3:14b      & 0.019 & 0.005  & 0.011 & 0.015   \\
 & command-r     & 0.633 & 0.547  & 0.223 & 0.563   \\
 & llama3.2      & 0.462 & 0.297  & 0.082 & 0.375   \\
 & nemotron-mini & 0.208 & 0.01   & 0.000 & 0.133   \\ \hline
\multirow{3}{*}{\begin{tabular}[c]{@{}l@{}}LLM w/ \\ itself as \\ as router\end{tabular}}                     & Gemma2:27b    & 0.598 & 0.59  & 0.341 & 0.566 \\
 & Mistral-small & 0.684 & 0.382  & 0.235 & 0.554   \\
 & Command-r     & 0.621 & 0.505  & 0.247 & 0.547   \\ \hline
\end{tabular}%
}
\end{table}

\subsection{Backward Inference Thinking}

\begin{figure}
    %\centering
    % \hspace{-.6cm}
    \includegraphics[width=1\linewidth]{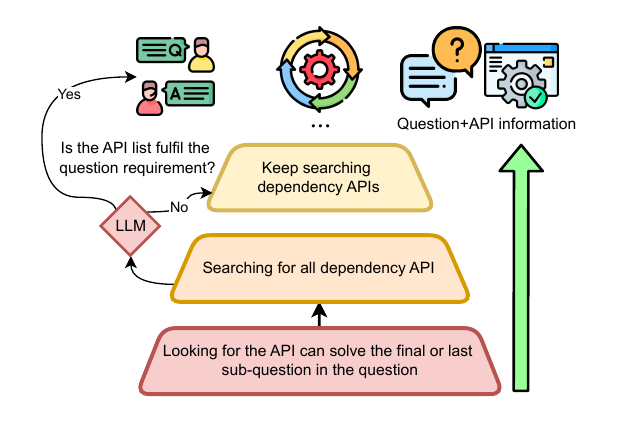}
    \caption{Backward Thinking Pipeline}
    \label{fig:backward}
    \vspace{-1.5em}
\end{figure}

To optimize the API selection and calling process, we implement a \textbf{Backward Thinking} approach, inspired by CauseJudger~\cite{he2024causejudger} and Reverse Chain~\cite{zhang2023reverse}, as illustrated in Figure~\ref{fig:backward}. This approach enables the model to construct a sequence of API calls more systematically by working backwards from the final goal rather than following a purely forward selection strategy.

The process follows these steps:

\begin{enumerate}
    \item \textbf{Identifying the Final API Call}: The model first determines the ultimate API needed to answer the user’s query. This API must provide the final required information or action.
    
    \item \textbf{Checking Parameter Completeness}: The model verifies whether all required parameters for the final API are available. If any essential information is missing, the model does not proceed with execution but instead considers the necessary steps to obtain the missing data.

    \item \textbf{Determining Supporting API Calls}: If missing parameters are identified, the model searches for additional APIs that can retrieve the necessary data. These supporting API calls are planned in reverse order, ensuring that the final API call has all the required inputs.

    \item \textbf{Iterative Refinement}: This process continues iteratively. Each newly identified API is analyzed for its own dependencies, ensuring that all required information is recursively retrieved before execution.

\end{enumerate}

By breaking the task into smaller, dependency-aware steps, this method allows the model to effectively plan and execute multi-step API calls, improving accuracy in complex scenarios. As shown in Table~\ref{tb:backward}, this approach yields a 30\% improvement in hard-level API calling tasks. The backward inference mechanism significantly enhances the model’s ability to handle intricate, real-world API calling scenarios, reducing failure cases caused by missing or misordered API dependencies.
\begin{table}[]
\caption{Backward Thinking performance in High difficulty calling in GPT-4o and GPT-4o-mini.}
\label{tb:backward}
\resizebox{\columnwidth}{!}{%
\begin{tabular}{lcccc}
\hline
\multirow{2}{*}{Model} & \multicolumn{2}{l}{API Calling Routing}                                & \multicolumn{2}{l}{API Calling with Parameters}                        \\ \cline{2-5} 
                       & Original & \begin{tabular}[c]{@{}l@{}}Backward\\ Thinking\end{tabular} & Original & \begin{tabular}[c]{@{}l@{}}Backward\\ Thinking\end{tabular} \\ \hline
GPT4o      & 0.611 & 0.894 & 0.388 & 0.729 \\
GPT4o mini & 0.564 & 0.847 & 0.364 & 0.482    \\ \hline
\end{tabular}%
}
\vspace{-2em}
\end{table}

\begin{tcolorbox}[leftrule=0mm,rightrule=0mm,toprule=0mm,bottomrule=0mm,left=0pt,right=0pt,top=0pt,bottom=0pt,title={RQ3:How to enhance API function calling routing ability for zero/few-shot LLM?}]
    \textbf{Answer: } We have tested 2 different ways to enhance the performance of function calling in our test, both can significantly increase performance in function calling routing and JSON generation.
\end{tcolorbox}
\section{Case Study}
\subsection{Insufficient Context Limit}
We observe that models like xLAM and NemoTron-Mini, which have a 4K context limit, struggle with longer API calls in CallNavi, where some inputs exceed 6K tokens. This limitation leads to truncated inputs, causing incorrect API selection and missing parameters in multi-call sequences. While models with higher context limits generally perform better, we also find that context length alone does not guarantee success—models must still effectively manage dependencies and navigate complex API workflows. These findings highlight the need for both expanded context windows and improved structured reasoning in function-calling tasks.
\subsection{Hallucination}
In one of our test cases, the \textbf{Phi3:14b} model produced an incorrect API function call in response to a baggage tracking scenario. The predicted output was as follows:

\begin{lstlisting}[language=json,firstnumber=1]
{
    "API": ["getLostBaggageReport", "updateBaggageStatus"],
    "parameters": [{"baggageId": "BAG123"}, {}]
}
\end{lstlisting}

However, the ground truth was:

\begin{lstlisting}[language=json,firstnumber=1]
{
    "API": ["getBaggageStatus"],
    "parameters": [{"baggageId": "BAG123"}]
}
\end{lstlisting}

In this case, the model hallucinated two API calls, \textbf{"getLostBaggageReport"} and \textbf{"updateBaggageStatus"}, which were not part of the provided API list. This hallucination led the model to predict incorrect API calls, deviating from the expected function \textbf{"getBaggageStatus"}. Although the model correctly captured the parameter \textbf{baggageId: "BAG123"}, it introduced an unnecessary second parameter block as an empty dictionary, further reducing the accuracy of the output. 

This example highlights a common issue with current large language models in complex tasks: their tendency to hallucinate irrelevant API calls when uncertain. Such behavior emphasizes the need for improved mechanisms to ensure more accurate API function routing and parameter generation in these models.
\subsection{JSON Generation}
In another example, the \textbf{Mistral-Nemo} model generated an incorrect output, which included unwanted notes in the result, rendering it invalid as a JSON. The predicted output was:

\begin{lstlisting}[language=json,firstnumber=1]
{
    "API": ["getCustomerCreditCards"],
    "parameters": [{"customerID": "123456"}]
}
#(Assuming that ATM cards are considered credit cards for this specific API)
\end{lstlisting}

The ground truth, however, was:

\begin{lstlisting}[language=json,firstnumber=1]
{
    "API": ["getATMCardList"],
    "parameters": [{"accountID": "123456"}]
}
\end{lstlisting}

In this case, the model incorrectly generated an API call for \textbf{"getCustomerCreditCards"} instead of the correct API \textbf{"getATMCardList"}. Additionally, the model included an unwanted note— \textbf{"(Assuming that ATM cards are considered credit cards for this specific API)"}—which made the output non-compliant with JSON formatting, as this additional text was outside the structure of the JSON object. 

This example illustrates the challenge of maintaining output fidelity in models when they generate explanations or assumptions within the response, which should be avoided in strict JSON-formatted outputs. Such behavior disrupts the automation of API calls and highlights the need for better prompt engineering to ensure models only return valid JSON results without extraneous content.
\subsection{Logical Errors in Hard Questions}
Logical errors are particularly prevalent in hard questions, where the task involves multiple dependent API calls or complex reasoning. These errors include incorrect sequencing of API calls, failure to propagate parameters correctly, or omitting necessary steps. e.g.:
\begin{itemize}
    \item \textbf{Example}: When asked to retrieve a user's transaction history and compute their monthly spending, the model retrieves the transactions but fails to invoke an API for computation, leaving the task incomplete.
    \item \textbf{Impact}: Logical errors highlight the limitations of current models in handling multistep tasks' dependency reasoning.
\end{itemize}
\subsection{Impact of JSON and YAML on Model Performance}
To analyze the influence of input and output formats on model performance, we conducted experiments using JSON and YAML, two widely used structured data formats. These formats differ significantly in syntax and structure, which could affect the ability of models to interpret, process, and generate outputs accurately. We tested four configurations:
\begin{itemize}
    \item \textbf{YAML to YAML}: Both input and output are YAML.
    \item \textbf{JSON to JSON}: Both input and output are JSON.
    \item \textbf{YAML to JSON}: Input data is formatted in YAML, and output data is in JSON.
    \item \textbf{JSON to YAML}: Input data is formatted in JSON, and output data is in YAML.
\end{itemize}

\noindent
The results of these experiments are shown in Table~\ref{tab:yamljson}.

% \usepackage{multirow}
% \usepackage{graphicx}
\begin{table}[htb]
\caption{Compare JSON or YAML as input/output performance differences.}
\label{tab:yamljson}
\centering
\resizebox{0.5\textwidth}{!}{%
\begin{tabular}{llrrrrrrr}
\hline
\begin{tabular}[c]{@{}l@{}}Input/\\ Output\end{tabular} &
  Model &
  \multicolumn{1}{l}{Syntax Acc} &
  \multicolumn{1}{l}{Structure Acc} &
  \multicolumn{1}{l}{Easy} &
  \multicolumn{1}{l}{Medium} &
  \multicolumn{1}{l}{Hard} &
  \multicolumn{1}{l}{Overall} &
  \multicolumn{1}{l}{GPT-Score} \\ \hline
\multirow{4}{*}{\begin{tabular}[c]{@{}l@{}}YAML\\ to\\ YAML\end{tabular}} & LLAMA3.1      & 0.525 & 0.076 & 0.081 & 0.042 & 0.011 & 0.063 & 0.183 \\
                                                                          & mistral-small & 0     & 0.183 & 0.23  & 0.047 & 0.035 & 0.16  & 0.241 \\
                                                                          & Gemma2:27b    & 0.938 & 0.097 & 0.12  & 0.037 & 0     & 0.085 & 0.238 \\
                                                                          & command-r     & 0.883 & 0.096 & 0.105 & 0.042 & 0.011 & 0.078 & 0.241 \\ \hline
\multirow{4}{*}{\begin{tabular}[c]{@{}l@{}}JSON\\ to\\ JSON\end{tabular}} & LLAMA3.1      & 0.925 & 0.207 & 0.223 & 0.058 & 0.059 & 0.162 & 0.422 \\
                                                                          & mistral-small & 0.986 & 0.196 & 0.201 & 0.106 & 0.059 & 0.16  & 0.417 \\
                                                                          & Gemma2:27b    & 0.982 & 0.226 & 0.217 & 0.143 & 0.07  & 0.181 & 0.476 \\
                                                                          & command-r     & 0.969 & 0.189 & 0.167 & 0.095 & 0.047 & 0.134 & 0.4   \\ \hline
\multirow{4}{*}{\begin{tabular}[c]{@{}l@{}}YAML\\ to\\ JSON\end{tabular}} & LLAMA3.1      & 0.88  & 0.194 & 0.208 & 0.106 & 0.023 & 0.16  & 0.347 \\
                                                                          & mistral-small & 0.995 & 0.179 & 0.173 & 0.112 & 0.058 & 0.144 & 0.333 \\
                                                                          & Gemma2:27b    & 0.984 & 0.212 & 0.195 & 0.159 & 0.082 & 0.172 & 0.414 \\
                                                                          & command-r     & 0.967 & 0.198 & 0.168 & 0.122 & 0.071 & 0.145 & 0.322 \\ \hline
\multirow{4}{*}{\begin{tabular}[c]{@{}l@{}}JSON\\ to\\ YAML\end{tabular}} & LLAMA3.1      & 0.598 & 0.104 & 0.14  & 0.026 & 0     & 0.094 & 0.32  \\
                                                                          & mistral-small & 0     & 0.218 & 0.263 & 0.079 & 0     & 0.185 & 0.332 \\
                                                                          & Gemma2:27b    & 0.931 & 0.128 & 0.153 & 0.026 & 0     & 0.103 & 0.419 \\
                                                                          & command-r     & 0.853 & 0.091 & 0.123 & 0.011 & 0     & 0.079 & 0.363 \\ \hline
\end{tabular}%
}
\vspace{-1em}
\end{table}

\subsubsection*{Analysis of Results}
\textbf{1. JSON Outperforms YAML.}  
Across all configurations, models achieved higher syntax, structure, and task-specific accuracies with JSON as both the input and output format. For example, the \textbf{JSON to JSON} configuration resulted in the highest Syntax Accuracy (e.g., 0.986 for Mistral-Small) and Structure Accuracy (e.g., 0.226 for Gemma2:27B), highlighting JSON's straightforward syntax and reduced ambiguity.

\textbf{2. YAML Challenges.}  
Models struggled significantly with YAML, particularly in the \textbf{YAML to YAML} configuration, which had the lowest performance across metrics. For instance, LLAMA 3.1 achieved a Syntax Accuracy of 0.525, and Structure Accuracy remained poor across models. YAML's indentation-sensitive syntax and verbosity likely contribute to these challenges.

\textbf{3. Mixed Configurations Mitigate Errors.}  
Configurations with mixed input and output formats (e.g., \textbf{YAML to JSON}) performed better than pure YAML setups. JSON as an output format simplified generation tasks, as evidenced by improved metrics compared to YAML outputs.

\textbf{4. JSON to YAML is Challenging.}  
The \textbf{JSON to YAML} configuration showed decreased performance compared to \textbf{JSON to JSON}, particularly in Syntax Accuracy (e.g., 0.598 for LLAMA 3.1). This indicates that YAML's complexity as an output format negatively affects model performance.
\section{Discussion and Conclusion}
\subsection{Discussion}
Our dataset introduces significant challenges, particularly in \textbf{medium} and \textbf{hard} questions, where models must select APIs from a large pool and generate parameters in \textbf{multi-step} and \textbf{nested contexts}. This complexity highlights the limitations of fine-tuned models trained on smaller API sets and underscores the need for more diverse and robust training paradigms.

We observe distinct model behaviours in API routing and parameter JSON generation. \textbf{GPT-4o} excels in both tasks, while models like \textbf{LLaMA 3.1} and \textbf{Gemma2} perform well in API routing but struggle with parameter generation, making them suitable for routing-centric applications. In contrast, smaller models (\textless 10B parameters) exhibit instability, often producing inconsistent or incomplete outputs, limiting their effectiveness in complex, multi-step scenarios.

\textbf{Long-context processing} remains a significant bottleneck. Although models with larger context windows better handle structured inputs, they still struggle with simultaneous logical inference and structured JSON generation. Our findings suggest that merely increasing context size does not fully resolve multi-step reasoning challenges, emphasizing the need for improved architectures and reasoning strategies such as \textbf{Chain of Thought}.

Despite advancements, no model, including \textbf{GPT-4o}, fully solves intricate API calling tasks, reinforcing the need for further research in LLM-driven function calling.

\subsection{Conclusion}
This work introduces \textbf{\texttt{CallNavi}}, a benchmark evaluating API function calling in LLMs across \textbf{500 APIs} and \textbf{700 questions}. We assess general-purpose and fine-tuned models, revealing key limitations in \textbf{API selection, parameter generation, and multi-step reasoning}. 

To improve function calling accuracy, we propose \textbf{2-steps generation} and \textbf{backward inference}, enhancing structured API selection. While larger models like \textbf{GPT-4o} perform well, they still struggle with \textbf{long-context input processing}, particularly in tasks requiring both \textbf{logical inference and structured JSON generation}. Models with \textless 6K token limits often truncate inputs, leading to incomplete API calls and degraded performance.

Our findings contribute to the broader field of \textbf{software engineering evaluation and assessment}, particularly in automated API function that calls for AI-based software design, stability evaluation and structured reasoning. Future work should focus on improving LLM robustness in real-world deployments, integrating \textbf{retrieval-augmented techniques}, and expanding function-calling benchmarks to incorporate real-time constraints such as \textbf{error handling, authentication, and API versioning}.

\section*{Threats To Validity}
\paragraph{Internal Validity}
One key limitation is context length constraints, where models like xLAM and NemoTron(4K) struggle with inputs exceeding 6K tokens in CallNavi, leading to truncation and incomplete API calls. While models with longer context windows perform better, our results suggest that context size alone is insufficient without strong reasoning and structured generation capabilities.

The complexity and variability of CallNavi, particularly in multi-step and nested API tasks, pose additional challenges. Fine-tuned models, often trained on smaller API sets, may struggle to generalize. Additionally, LLM-as-a-judge introduces potential subjectivity in evaluation. Our optimization strategies like 2-steps generation and backward inference—improve multi-step API selection, but their effectiveness may vary across different architectures.

\paragraph{External Validity}
While CallNavi spans 500+ APIs and 700+ questions across 10 domains, it does not cover all real-world constraints, such as authentication, error handling, or evolving API versions. Future LLMS with longer context, hybrid architectures, etc., may demonstrate different performance trends.

Additionally, real-world API integration involves challenges beyond our benchmark, such as network failures, rate limits, and dynamic tool adaptation. While our evaluation covers syntax validity, AST match, and stability, future extensions should explore live production-level testing to better assess real-world deployment challenges.

\section*{Acknowledgment}
We thank our collaborator BGL BNP Paribas for their support. The FNR funded this research under grants NCER22/IS/16570468/NCERFT and BRIDGES2021/IS/16229163/LuxemBERT.

% \begin{acks}
% \end{acks}


\bibliographystyle{ACM-Reference-Format}
\bibliography{sample-base}

% \appendix

% \input{appendix}

\end{document}